\documentclass[aps,prl,twocolumn,groupedaddress,showpacs]{revtex4}
\usepackage[pdftex]{graphics}

\begin{document}

\title{Weak Ferromagnetic Exchange and Anomalous Specific Heat in ZnCu$_3$(OH)$_6$Cl$_2$}

\author{Ookie Ma}
\email[]{Seungwook_Ma@brown.edu}
\author{J. B. Marston}
\email[]{Brad_Marston@brown.edu}
\affiliation{Department of Physics, Brown University, Providence, RI USA 02912-1843}

\date{\today}

\begin{abstract}
Experimental evidence for a plethora of low energy spin excitations in the spin-1/2 kagome antiferromagnet ZnCu$_3$(OH)$_6$Cl$_2$ may be understandable in terms of an extended Fermi surface of spinons coupled to a U(1) gauge field.   We carry out variational calculations to examine the possibility that such a state may be energetically viable.  A Gutzwiller-projected wavefunction reproduces the dimerization of a kagome strip found previously by DMRG.  Application to the full kagome lattice shows that the inclusion of a small ferromagnetic next-nearest-neighbor interaction favors a ground state with a spinon Fermi surface. 

\end{abstract}


\maketitle

The spin-1/2 Heisenberg antiferromagnet on a kagome lattice is a proving ground for the existence of a two-dimensional spin liquid.  Can the combination of low spin, low coordination number, and geometric frustration lead to spin disordered ground states and excitations with fractional statistics?  If so, then it remains to establish whether spin-spin correlations are short-ranged with a gap to triplet excitations or whether the correlations are quasi-long-ranged with gapless excitations.  Experiment is the ultimate arbiter of these fundamental questions.  Thus, the recent experimental realization of a spin-1/2 kagome antiferromagnet (KAF), ZnCu$_{3}$(OH)$_6$Cl$_2$ also known as Herbertsmithite\cite{Shores2005}, is of great interest.   Magnetic susceptibility, muon spin rotation, and NMR measurements show no evidence of magnetic order\cite{helton-2007prl, imai-2007}.  The bulk spin susceptibility exhibits Curie behavior down to temperature $T=0.5J$, where $J \approx$ 170 K, but then increases sharply with decreasing temperature, possibly saturating at $T=0$.  Likewise, the specific heat shows significant enhancement at low temperatures.  At very low temperatures, the temperature dependence can be roughly fit to a power law with an exponent as small as $\alpha=1/2$.  Over the range $T = 5$ to  $30 \times 10^{-4}J$, a best fit\cite{ofer-2006} to $C_v\propto T^\alpha$ yields $\alpha = 2/3$.  A small magnetic field suppresses the enhancement, suggesting that the contributing excitations are magnetic in origin. 

Various explanations for the experiments have been proposed.  For instance, it has been observed that samples of Herbertsmithite contain a significant number of impurties.  Saturation of the susceptibility may be a Kondo effect due to impurity substitution of non-magnetic zinc ions by copper atoms\cite{ran-2007}.  A separate contribution from isolated spinful inpurities has been argued to explain the measured bulk susceptibility\cite{misguich-2007}.  Furthermore, the temperature variation of the specific heat in an applied magnetic field resembles a Schotty anomaly and is similar to that attributed to defects in the spin-1 chains Y$_2$BaNiO$_5$ and NENP\cite{ramirez, vries-2007}.  Powder NMR measurements of the local spin susceptibility suggest a second possibility.  Although the averaged local susceptibility tracks the bulk, regions of the NMR spectrum actually show a decrease in the susceptibility with decreasing temperature.  The discrepancy between local and bulk measurements may indicate the presence of impurities or may instead be attributed to the sampling of the susceptibility along different and distinct crystallographic directions\cite{imai-2007, olariu}.  If the latter case, the low temperature ferromagnetism may be due to a Dyzaloshinskii-Moriya (DM) anisotropy\cite{rigol:207204, rigolSinghM}.  

To make a clear prediction for the specific heat at the lowest experimental temperatures, we use Gutzwiller-projected mean-field theory.  Our approach is motivated in part by the appearance of a specific heat exponent $\alpha < 1$ found for an extended Fermi surface of spinons coupled to a U(1) gauge field\cite{lee-1992}.  A related composite-fermion theory has had some success in explaining the behavior of the half-filled Landau level\cite{halperin-1993}.   Once the fermions are integrated out, the contribution of the gauge field to the specific heat can be obtained from the free energy\cite{lee-1992, kim-1996}.  In the absence of long-ranged density-density interactions between the fermions, it takes the form
\begin{equation}
C(T) \propto T^{2/3}
\label{specificHeatExponent}
\end{equation}
and thus, dominates conventional contributions such as that due to phonons.  This signature provides motivation to investigate candidate states that possess large spinon Fermi surfaces.

In the case of the nearest-neighbor (NN) Heisenberg model, our calculations reproduce the results of Ref. \onlinecite{ran-2007}, namely, we find that the lowest energy state has no broken symmetries and the corresponding mean-field state exhibits Dirac fermions at nodal points.  Under the hypothesis that the experimentally observed weak ferromagnetism is intrinsic to the pure lattice of Cu$^{2+}$ spins, we investigate the effect of adding a small next-nearest-neighbor (NNN) interaction with negative (ferromagnetic) coupling
\begin{equation}
H = J_1 \sum_{\langle i,j \rangle} \mathbf{S}_i\cdot\mathbf{S}_j - J_2 \sum_{\langle\langle i,j \rangle\rangle} \mathbf{S}_i\cdot\mathbf{S}_j\ .
\label{hamiltonian}
\end{equation}
For convenience, we set $J_1 = 1$ in the following.   We note that establishing the sign of the NNN spin-spin interaction from first principles requires a detailed understanding of the quantum chemistry of ZnCu$_{3}$(OH)$_6$Cl$_2$.  An organic spin-1/2 KAF has been argued to have mixed ferromagnetic NN and antiferromagnetic NNN bonds\cite{narumi-2004}, but we leave open the possibility that other sub-leading terms, like the DM anisotropy, may also be important in a realistic model of Herbertsmithite. 

A rational approach to the systematic construction of variational spin liquid states can be realized by generalizing the ordinary SU(2) Heisenberg antiferromagnet to SU(N) spins living in the self-conjugate representation, meaning that each site has N/2 fermions\cite{marstonaffleck}.  The resulting model may then be solved exactly in the large-N limit by decoupling the spin-spin interaction with the Hubbard-Stratonovich field $\chi_{ij}=\langle f^{\dagger\alpha}_{i}f_{j\alpha}\rangle$.   Appropriate variational states are then obtained by the usual Gutzwiller-projection procedure\cite{arun}.   To compare candidate variational wavefunctions in the physical SU(2) limit, NN and NNN correlations are sampled by a standard determinantal wavefunction Monte Carlo algorithm\cite{gros, ceperly} that exactly enforces the single occupancy constraint $f^{\dagger\uparrow}_if_{i\uparrow}+f^{\dagger\downarrow}_if_{i \downarrow}=1$.  The $\chi_{ij}$ field, interpreted as a bond amplitude order parameter, may then be adjusted to optimize the ground state energy to test for various dimerization instabilities.  This approach has been shown to reproduce qualitatively the phase diagrams of certain spin chains that are known exactly by other means\cite{arun}.   

\begin{figure}
\resizebox{6.5cm}{!}{\includegraphics{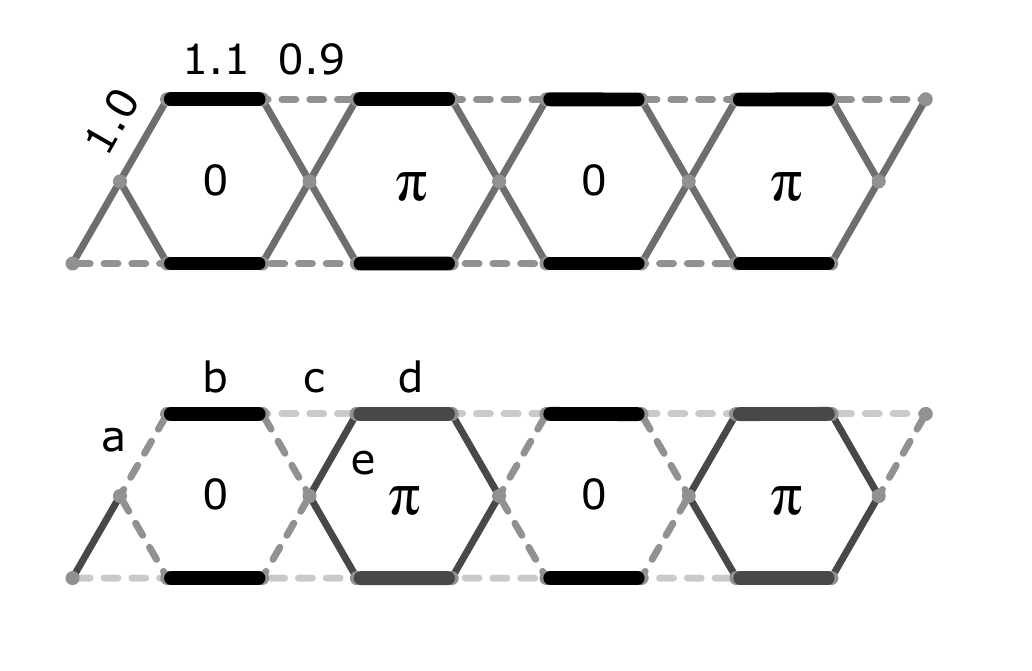}}
\caption{\label{stripDimer} The kagome strip.  Top: The optimal variational state.  Numerical values indicate the relative hopping magnitudes along the links and the value of the flux through each hexagon.  There is no flux through the triangles.  Bottom: The magnitude of the five symmetry-unrelated nearest-neighbor spin-spin correlation functions $\langle \mathbf{S}_i \cdot \mathbf{S}_j \rangle$, labeled (a) through (e), is indicated by the thickness of the lines.  See text for the actual values.}
\end{figure}

To test the reliability of this approach, we first study the nearest-neighbor Heisenberg model on a 2-leg kagome stip.  Because of its one-dimensional character, the ground state has been accurately described by the essentially exact numerical density matrix renormalization group (DMRG) method where it was shown\cite{whiteSingh} to be strongly dimerized with five different values for the NN spin-spin correlations $\langle \mathbf{S}_i \cdot \mathbf{S}_j \rangle$.  The strip employed in the DMRG calculation has open boundaries that are terminated in such a way to remove triplet excitations from the low energy spectrum; nevertheless the nearest-neighbor spin-spin correlations are well converged in the interior.  The NN spin-spin correlations (a) through (e), as labelled in Fig. \ref{stripDimer}, are $-0.16429$, $-0.63506$, $-0.07087$, $-0.52862$ and $ -0.33122$ respectively\cite{privWhite}, and the ground state energy per site based upon these interior values is $E_0 = -0.45929$.

For the variational calculation we adopt periodic boundary conditions.  There are three distinct bonds per unit cell.  The possibility of spontaneous symmetry breaking is then tested\cite{arun} by examining nontrivial patterns of gauge flux threading through triangles and hexagons.  States that break time-reversal symmetry always appear to have high energy after projection; hence we follow Ref. \onlinecite{hastings} and restrict attention only to fluxes 0 and $\pi$.   NN correlations are sampled on a strip with length 100 sites and 250 total spins.  We find that among states that at most double the size of the unit cell, the lowest energy state has a flux of $\pi$ through every other hexagon and zero flux through the remaining hexagons and all triangles; thus, the variational state exhibits dimer order.   The optimal arrangement of fluxes and bond magnitudes is depicted in the top panel of Fig. \ref{stripDimer}.  Along the two legs, alternating bond magnitudes of approximately 1.1 and 0.9 yield the lowest energy, with the remaining third bond fixed at unit hopping amplitude.   The doubled unit cell supports five symmetry-inequivalent values of the NN spin-spin correlation function $\langle \mathbf{S}_i \cdot \mathbf{S}_j \rangle$ in agreement with the DMRG results.   Along bonds (a) through (e) the spin-spin correlation functions are $-0.08649(48)$, $-0.67460(40)$, $-0.06213(42)$, $-0.46876(30)$, and $-0.41077(36)$.  The pattern shows good qualitative agreement with the DMRG calculation: the five distinct spin-spin correlation functions appear in the same order by size, and the variational energy per site of $E_0 = -0.45239(3)$ is only $1.5 \%$ higher than the DMRG value.   Further relaxation of the energy towards the DMRG value may be expected if all five distinct bonds were to be adjusted in magnitude.  

\begin{figure}
\resizebox{7cm}{!}{\includegraphics{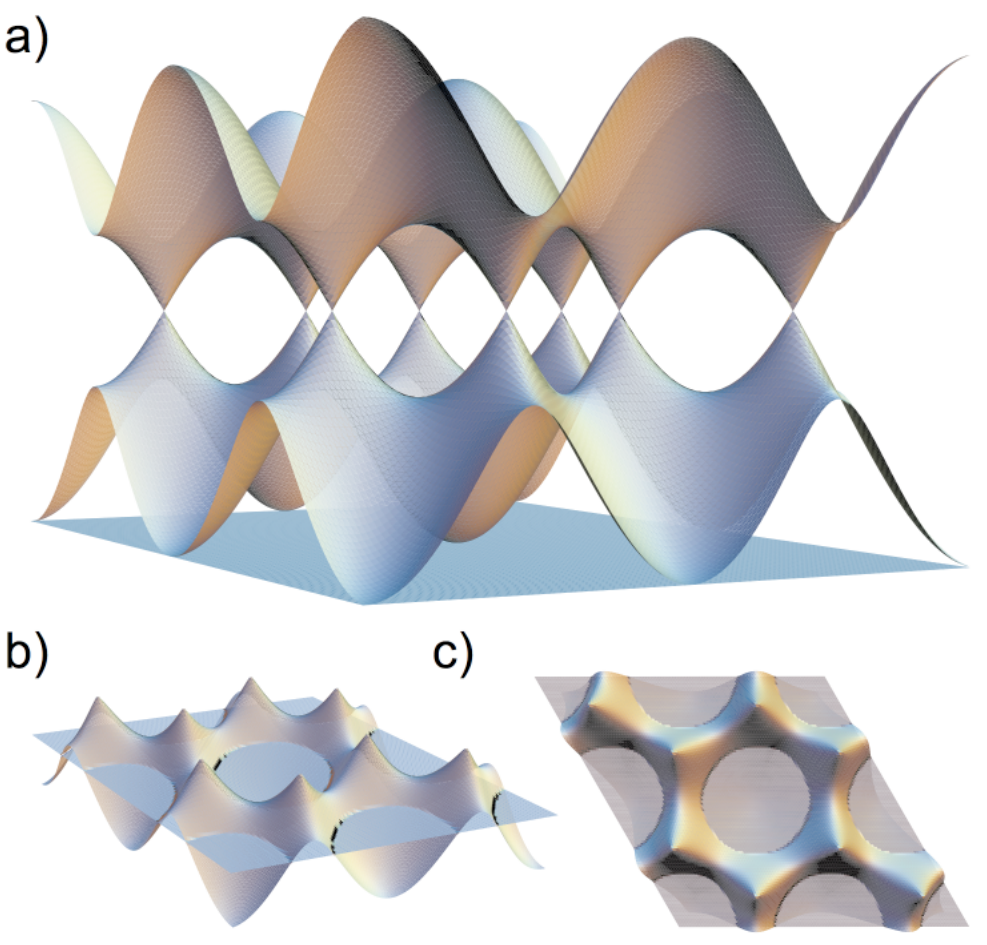}}
\caption{\label{uniform} (Color online) Mean-field spinon dispersion of the uniform phase of zero flux.   (a) Complete dispersion showing that one of the three bands is completely flat.  (b) Middle band cut by a plane at the Fermi energy corresponding to half-filling.  Note that the Dirac nodes are not located at the Fermi energy.  (c) Rotation of (b) to highlight the nearly circular Fermi surface.}
\label{spectrum}
\end{figure}

As the variational approach appears to give a good description of the kagome strip,  we now turn to the full 2D kagome lattice.  Several candidate spin liquid states have been examined\cite{ran-2007}.   One competitive state has flux $\pi$ through the hexagons and no flux through the triangles.  This $\pi$-hexagon state has Dirac nodes at the Fermi level\cite{ran-2007}.  Another state of interest has no flux penetrating through the lattice anywhere.  As shown in Fig. \ref{spectrum}, this uniform state possesses an extended Fermi surface at half-filling.  

The energies of the candidate states are calculated on an oblique lattice with $12 \times 12$ 3-site unit cells ($432$ total sites).  On finite lattices, the choice of boundary conditions is important.  In the case of periodic boundaries, there is an ambiguity in the filling of independent-particle states at the Fermi level, as only some of the degenerate independent-particle states are filled.  To resolve the ambiguity, the degeneracy must be lifted.  Mixed periodic and antiperiodic boundaries accomplishes this\cite{shiba} at the cost of breaking lattice rotational symmetry.  Consequently, the mixed boundary conditions induce modulation in observables such as $\langle \mathbf{S}_i \cdot \mathbf{S}_j \rangle$.  The modulation, however, decreases with increasing system size, vanishing in the thermodynamic limit.  For the relatively large lattice studied here, the modulation is smaller than the uncertainty in the calculated spin-spin correlation function due to the statistics of finite samples.  

\begin{figure}
\resizebox{6.5cm}{!}{\includegraphics{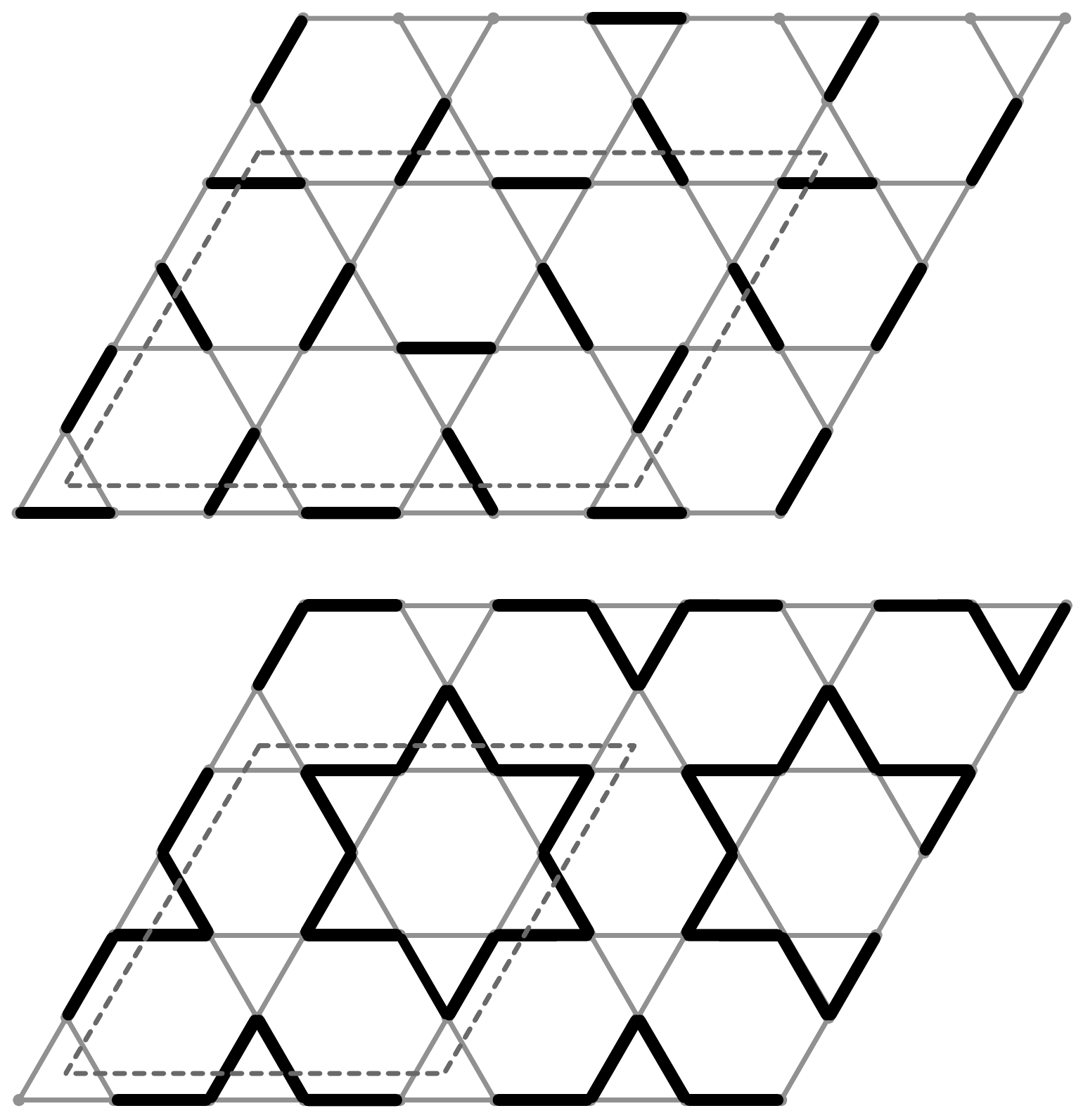}}
\caption{\label{dimer} Two possible dimer orderings with unit cells outlined by dashed parallelograms.  Thick lines indicate larger hopping magnitudes and hence, larger NN spin-spin correlations.  Top panel:  Dimer order that maximizes the number of hexagons with exactly three dimers\cite{marstonzeng}.  Bottom panel:  Plaquette order that outlines the shape of a star\cite{hastings}.}
\end{figure}

We check for instabilities of the variational states toward two different types of dimer patterns.  The first pattern is a dimer order of the type first proposed in Ref. \onlinecite{marstonzeng} based on $1/N$ corrections to the large-N solution.  Dimer coverings that maximize the number of hexagons with exactly three dimers (``perfect hexagons'') are energetically preferred because the perfect hexagons resonate like the alternating single and double bonds of a benzene ring.  As the resonance is a second-order process, it lowers the energy.  There can be at most one perfect hexagon for every 18 sites\cite{marstonzeng}.  The top panel in Fig. \ref{dimer} shows an 18-site unit cell with one perfect hexagon.  Larger unit cells support other possible dimer coverings that maximize the density of perfect hexagons including the honeycomb pattern shown in Fig. 1(b) of Ref. \onlinecite{marstonzeng} and the striped pattern shown in Fig. 5(b) of Ref. \onlinecite{senthil}.  For the NN Heisenberg model, a systematic dimer expansion determines that the honeycomb and stripe patterns have ground state energies per site of -0.432088216 and -0.4315321, respectively\cite{singhhuse}.  These energies compare quite well to the ground state energy of a 36 site cluster, -0.438377, found by exact diagonalization\cite{leungelser}.
The second dimer configuration has plaquette order of a type that has been proposed as a potential instability of the $\pi$-hexagon state.  As shown in the bottom panel of Fig. \ref{dimer}, the plaquette takes the shape of a star formed by the outer edges of six triangles that share a common hexagon.  This pattern of bonds would be induced by the dynamical formation of a non-chiral mass term that gaps out the nodal fermions\cite{hastings}.  

\begin{table}
\caption{\label{antiperiodic} The lattice-averaged spin-spin correlation function $\langle \mathbf{S}_i \cdot \mathbf{S}_j \rangle$ for an oblique lattice of  $12 \times 12$ 3-site unit cells (432 sites total) with mixed periodic and antiperiodic boundary conditions.  The magnitudes of the bond amplitudes are modulated by $5 \%$ for the ordered states.  Numbers in parentheses are the statistical errors due to the finite number of samples.}
\begin{ruledtabular}
\begin{tabular}{lcc}
& $( i,j ) \in$ NN & $( i,j ) \in$ NNN \\
$\pi$-hexagon state & -0.42866(2) & -0.02297(21)\\
$\qquad$dimer order & -0.42856(3) & -0.02277(22) \\
$\qquad$plaquette order & -0.42841(4) & -0.02271(17) \\
uniform state & -0.41215(2) & 0.07724(13)\\
$\qquad$dimer order & -0.41205(2) & 0.07873(10)\\
$\qquad$plaquette order & -0.41203(2) & 0.07701(13)\\
\end{tabular}
\end{ruledtabular}
\end{table}§

Table \ref{antiperiodic} summarizes the variational results.  Instabilities are tested by imposition of a small ($5 \%$) modulation in the magnitude of the $\chi_{ij}$ bond amplitudes.  In addition to the tabulated results, we also find that imposition of a $5 \%$ modulation of the honeycomb type upon the $\pi$-hexagon state raises the energy only slightly, to -0.42860(3), while at the same time gapping out the Dirac nodes.  This result lends support to the idea\cite{marstonzeng,senthil,singhhuse} that the nearest-neighbor antiferromagnet is actually dimer ordered.    Furthermore, all of the dimer coverings we tested that maximize the number of perfect hexagons are energetically preferable to the star plaquette pattern\cite{hastings,ran-2007}.  We speculate that further variational tuning will yield an ordered state of perfect hexagons that is energetically competitive with the best estimates for the ground state energy of the NN spin-$1/2$ KAF.

Because the NNN spin-spin correlations are negative in the $\pi$-hexagon phase but positive in the uniform phase, the uniform state is energetically favored over the $\pi$-hexagon state for sufficiently large ferromagnetic NNN coupling, $J_2 > 0.16$ (the ground state remains antiferromagnetic up to much larger values of $J_2$).  At $J_2$ = 0.16, dimer modulation of the uniform state further lowers the energy, with a minimum in the energy occurring at approximately $4 \%$ bond modulation.  This $4 \%$ bond modulation induces a $18 \%$ modulation in the value of $\langle \mathbf{S}_i\cdot \mathbf{S}_j\rangle$ on symmetry-distinct bonds.  Somewhat surprisingly, the mean-field theory continues to support gapless excitations despite the broken translational and rotational symmetries, with little change in the density of states at the Fermi energy.  This is because ordering wavevectors corresponding to large (multiple of 18 sites) unit cells do not span the spinon Fermi surface.   Thus, a small ferromagnetic NNN interaction favors a spin-liquid phase with a large spinon Fermi surface, suggesting that spinons can remain deconfined even in the presence of dimer order.  We are not aware of any rigorous arguments (eg. Ref. \onlinecite{hastings2004}) that preclude this possibility.  

The large spinon Fermi surface exhibits at low temperature both a relatively large (but finite) magnetic susceptibility and an enhanced specific heat, Eq. \ref{specificHeatExponent}, due to interactions between spinons and the U(1) gauge field.  These low energy properties of the uniform state agree qualitatively with those from available Herbertsmithite experiments, though to reproduce quantitatively the anomalously large susceptibility, it may be necessary to invoke additional contributions from impurities\cite{misguich-2007} and/or the DM interaction\cite{rigol:207204, rigolSinghM}. 

\section{Acknowledgments}

We thank A. Paramekanti for technical assistance with the determinant update algorithm; S. White for providing us with DMRG results; and J. Fj{\ae}restad, M. Hermele, R. McKenzie, V. Mitrovi{\'c}, R. Singh, and M.-A. Vachon for helpful discussions.  Some of the work was carried out at the 2007 summer workshop on ``Novel Aspects of Superconductivity'' at the Aspen Center for Physics.  It was supported in part by the National Science Foundation under grant No. DMR-0605619.  

\bibliography{paper.bib}

\end{document}